\documentclass[twocolumn,showpacs, showkeys,preprintnumbers,amsmath,amssymb, prb]{revtex4}
\usepackage{graphicx}
\usepackage{dcolumn}
\usepackage{bm}
\usepackage[dvips]{color}
\begin{document}

\preprint{Accepted in Phys. Rev. B}

\title{Anisotropic magnetic and superconducting properties of \\ CaFe$_{2-x}$Co$_x$As$_2$ ($x$ = 0, 0.06) single crystals}
\author{Neeraj Kumar, R. Nagalakshmi, R. Kulkarni, P. L. Paulose, A. K. Nigam, S. K. Dhar and A. Thamizhavel}
\affiliation{Department of Condensed Matter Physics and Material
Sciences, Tata Institute of Fundamental Research, Homi Bhabha Road,
Colaba, Mumbai 400 005, India.}
\date{\today}

\begin{abstract}
We report anisotropic dc magnetic susceptibility $\chi(T)$, electrical resistivity $\rho(T)$, and heat capacity $C(T)$ measurements on the single crystals of CaFe$_{2-x}$Co$_x$As$_2$ for $x$ = 0 and 0.06. Large sized single crystals were grown by the high temperature solution method with Sn as the solvent. The SDW/structural transition observed at 170~K in the pure CaFe$_2$As$_2$ compound is suppressed for the Co-doped sample CaFe$_{1.94}$Co$_{0.06}$As$_2$ and superconductivity is observed at $\simeq$~17~K.  The superconducting transition has been confirmed from the magnetization and electrical resistivity studies.  The $^{57}$Fe M\"{o}ssbauer spectrum in CaFe$_2$As$_2$ indicates that the SDW ordering is incommensurate with the crystal lattice.  In the Co-doped sample, a prominent paramagnetic line at 4.2~K is observed indicating a weakening of the SDW state.

\end{abstract}



\maketitle

The discovery of superconductivity at a high transition temperature $T_{\rm c}$~$\approx$~26~K in iron based LaFeAsO$_{1-x}$F$_x$~\cite{Kamihara} has created a torrent of research activity in the field of magnetism and superconductivity.  The $T_{\rm c}$ is significantly enhanced in these oxypnictides both by the application of external hydrostatic pressure~\cite{Takahashi} ($T_{\rm c}~\approx~$43~K at 4~GPa) and replacing La by smaller sized rare earth ions ($T_{\rm c}~\approx$~43~K in the iso-structural compound SmFeAsO$_{1-x}$F$_x$~\cite{Chen}).  The crystal structure of these compounds is characterized by an alternative stacking of LaO and FeAs  layers like LaO--FeAs--LaO--FeAs, where the Fe and As atoms are strongly coupled by  covalent bond~\cite{Takahashi}.  Although the oxypnictides possess high superconducting transition temperature crystal growth of a sizeable amount has not been achieved till date, to understand the anisotropic physical properties and the nature of superconductivity.  More recently, another class of FeAs based compounds AFe$_2$As$_2$ (A = Ba, Ca, Sr and Eu)~\cite{Rotter, Yan, Ren, Ronning} have been identified in which superconductivity can be induced either by doping or by applying hydrostatic pressure.  For example, K (hole) doped Ba$_{0.55}$K$_{0.45}$Fe$_2$As$_2$ shows a $T_{\rm c}$ as high as 38~K~\cite{Goldman} while Co (electron) doped BaFe$_{1.8}$Co$_{0.2}$As$_2$ exhibits a superconducting transition at 22~K~\cite{Sefat}.  The crystal structure of AFe$_2$As$_2$ contains FeAs layers, similar to oxypnictides, which are separated by the layers of A atoms.    Unlike the oxypnictides, the AFe$_2$As$_2$ compounds can be grown in single crystalline form, by high temperature solution growth method using either a Fe-As self flux taking advantage of the Fe-As binary eutectic or with  Sn as flux.   

While new results are being reported at a fast pace, to our knowledge the results of Co doped single crystalline CaFe$_2$As$_2$ have not been reported till date.  In this communication, we present our results on the anisotropic properties of pure and Co-doped single crystals of CaFe$_2$As$_2$.
Single crystals of CaFe$_{2-x}$Co$_{x}$As$_2$ ($x$ = 0 and 0.06)  were grown by the high temperature solution growth using Sn as flux. The starting materials  were high pure metals of  Ca, (Fe,Co), As and Sn taken in the ratio 1 : 2 : 2 : 19 for pure compound;  for the Co-substituted sample our starting composition was 1 : 1.8 : 0.2 : 2 : 19.  The contents were placed in a high quality recrystallized alumina crucible, and subsequently sealed in an evacuated quartz ampoule.  The furnace was slowly heated to 1050~$^\circ$C and held at that temperature for 24 hours to ensure homogenization. It was gradually cooled down to 450~$^\circ$C over a period of 3 weeks and then rapidly brought down to room temperature.  The flux was removed by means of centrifuging.  Very large sized single crystals with the typical dimensions of $20~\times~12~\times~0.4~$~mm$^3$, were obtained.  Laue patterns showed the $c$-axis is normal to the flat plane of the crystal.  The crystals were silvery white and malleable.  Electron probe micro-analysis showed that the composition of the Co-doped single crystal is CaFe$_{1.94}$Co$_{0.06}$As$_2$.  

The phase purity of the single crystalline sample was confirmed by performing powder x-ray diffraction by grinding a few pieces of the single crystals. The x-ray pattern can be indexed to the ThCr$_2$Si$_2$ type body centered tetragonal structure with the space group $I4/mmm$. The estimated lattice parameters are $a$~=~3.8942(8)~\AA~and $c$~=~11.746(7)~\AA~ for CaFe$_2$As$_2$ and $a$~=~3.8875(1)~\AA~and $c$~=~11.687(1)~\AA~ for CaFe$_{1.94}$Co$_{0.06}$As$_2$. The lattice parameter of the pure sample is in agreement with the previously reported values~\cite{Ronning, Goldman}.  The Co-doping compresses the $c$ axis by 0.5\% and the $a$~axis by 0.17\%. 

\begin{figure}[h]
\includegraphics[width=0.35\textwidth]{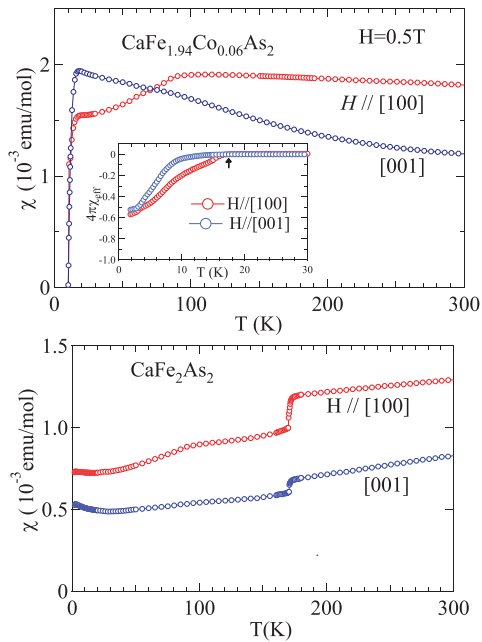}
\caption{\label{fig1}(Color online) (a)  Temperature dependence of magnetic susceptibility of CaFe$_{1.94}$Co$_{0.06}$As$_2$ in a field of 0.5~T, the inset shows the low temperature part of the dc magnetic susceptibility measured in 0.005~T, the arrow at 17~K indicates the onset of diamagnetic susceptibility, (b) Temperature dependence of the magnetic susceptibility  of  CaFe$_2$As$_2$ for $H$~$\parallel$~[100] and [001] in an applied magnetic field of 5~T.  }
\end{figure}

Figure~\ref{fig1}(a) shows the temperature dependence of the magnetic susceptibility of CaFe$_{1.94}$Co$_{0.06}$As$_2$ measured in a field of 0.5~T.  For the sake of comparison the corresponding data of CaFe$_2$As$_2$ in an applied field of 5~T is shown in Fig.~\ref{fig1}(b).   A sharp drop (Fig.~\ref{fig1}(b)) in the magnetic susceptibility at 170~K is a signature of a SDW magnetic ordering associated with the iron moments and a lattice structural transition in CaFe$_2$As$_2$. In this regard our data are in agreement with earlier reports in the literature  ~\cite{Ronning, Ni1, Wu}. In particular the magnitude of {$\chi$} for $H~\parallel~$[001] is nearly the same as in  Ni \textit{et al.}~\cite{Ni1}. The 170 K anomaly has vanished for the Co-doped sample indicating a suppression of the SDW ordering.  However, the susceptibility of CaFe$_{1.94}$Co$_{0.06}$As$_2$ along [100] shows an anomaly near 90~K  which may arise due to an antiferromagnetic like ordering of the residual Fe moments.   Incidentally we see a similar feature along [100] around the same temperature in CaFe$_2$As$_2$, but not observed by others  ~\cite{Ronning, Ni1, Wu}, which may imply a re-orientation of the Fe moments.  At lower temperatures  a sharp drop in the susceptibility near 17~K in CaFe$_{1.94}$Co$_{0.06}$As$_2$,  attaining diamagnetic values with the decrease of temperature is a signature of a superconducting transition. The inset of Fig.~\ref{fig1}(a)  shows the diamagnetic seceptibilty, corrected for demagnetization factor for $H~\parallel~[001]$~\cite{prozorov}, along both the principal directions measured in a field of 0.005~T. The superconducting volume fraction is found to be 55-60 \%.

\begin{figure}[h]
\includegraphics[width=0.35\textwidth]{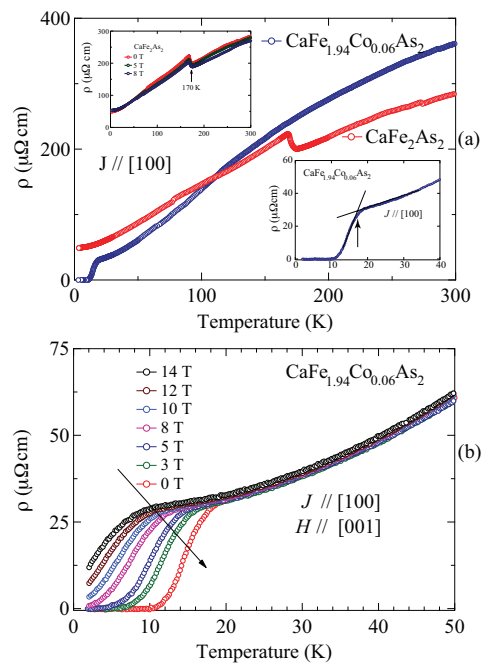}
\caption{\label{fig2}(Color online)(a) Temperature dependence of the electrical resistivity of CaFe$_2$As$_2$ and CaFe$_{1.94}$Co$_{0.06}$As$_2$ for J~$\parallel$~[100]; the upper inset shows the electrical resistivity of CaFe$_2$As$_2$ measured at 5 and 8~T and the lower inset shows the low temperature part of the superconducting CaFe$_{1.94}$Co$_{0.06}$As$_2$ sample.  The $T_{\rm c}$ in CaFe$_{1.94}$Co$_{0.06}$As$_2$ is found to be 17~K as shown in the lower inset.  (b) Resistivity measured at various applied magnetic fields of  CaFe$_{1.94}$Co$_{0.06}$As$_2$  for $J~\parallel$~[100] and $B~\parallel$~[001].}
\end{figure}

The temperature dependence of dc electrical resistivity  from 1.8 to 300 K for both pure and Co-doped CaFe$_2$As$_2$ single crystals for current parallel to [100] direction is shown in Fig.~\ref{fig2}(a).   The magnitude and the thermal variation of the electrical resistivity  of the pure compound are in agreement with the earlier reports~\cite{Ronning, Ni1}.  The sharp increase near 170~K is due to the SDW/structural transition.  The upper inset of  Fig.~\ref{fig2}(a) shows the electrical resistivity measured in 5 and 8~T for the pure CaFe$_2$As$_2$.  It is evident  that the magnetic field has virtually no effect on the electrical resistivity and that the SDW ordering is insensitive to the fields as high as 8~T. Our observation is in conformity with a similar conclusion arived earlier by Ni \textit{et al.}~\cite{Ni1} who found the first order phase transition at 170~K unaffected by fields as high as 14~T. On the other hand, the electrical resistivity of CaFe$_{1.94}$Co$_{0.06}$As$_2$ does not show any anomaly  at 170~K in accordance with the susceptibility data presented above.   It appears to vary smoothly across 90~K where the susceptibility along [100] shows an anomaly as discussed above.    At 17~K the electrical resistivity drops rapidly and becomes zero at lower temperature.  The lower inset of Fig.~\ref{fig2}(a) shows onset of superconducting transition on an expanded scale.  Figure~\ref{fig2}(b) shows the temperature dependence of the electrical resistivity below 50~K of CaFe$_{1.94}$Co$_{0.06}$As$_2$  for the current parallel to [100] in various applied magnetic fields ($H~\parallel~$[001]). Data with current and field both parallel to [100] were also recorded and showed a similar trend. As the field is increased the transition becomes broader and the $T_{\rm c}$ shifts to lower temperature  which is a characteristic feature of a type-II superconductor.  It may be noticed that we obtained a zero resistance state even in an applied field of 8~T, which indicates a high critical field $H_{\rm c2}$.  There is only a very marginal change in the normal state resistivity with applied field.  

\begin{figure}[h]
\includegraphics[width=0.35\textwidth]{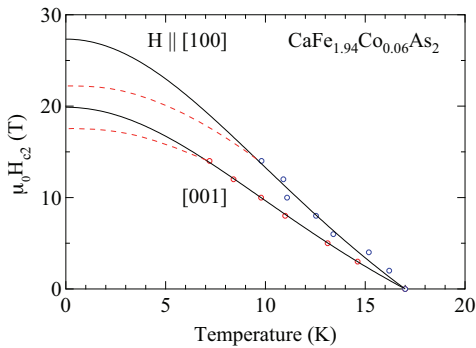}
\caption{\label{fig3}(Color online) Temperature dependence of the upper critical field $\mu_0 H_{\rm c2}(T)$ of CaFe$_{1.94}$Co$_{0.06}$As$_2$. The dashed line is the value of $\mu_0 H_{\rm c2}$ estimated by WHH theory and the solid line is according to Ginzburg-Landau theory (see text for details).}
\end{figure}
  
Using the procedure depicted in the lower inset of Fig.~\ref{fig2}(a) we can determine the temperature dependence of the upper critical field $H_{\rm c2}$ which is plotted in Fig.~\ref{fig3}.  The variation of $H_{\rm c2}$ with temperature is nearly linear with a negative slope and it does not show any kind of saturation for fields as high as 14~T. However one can estimate the $H_{\rm c2}(0)$ by using the Werthamer-Helfand-Hohenberg~\cite{WHH} (WHH) equation $H_{\rm c2}(0)=-0.7T_{\rm c}(dH_{\rm c2}/dT_{\rm c})$. The slope $dH_{\rm c2}/dT_{\rm c}$ is estimated to be -1.82~T/K for $H~\parallel$~[100] and $-1.43$~T/K for $H~\parallel~$[001].  For a $T_{\rm c}$ of 17~K, the upper critical field $H_{\rm c2}(0)$ is found to be 22~T and 17~T, for $H~\parallel~$ [100] and [001] directions respectively,  as shown by the dotted line in Fig.~\ref{fig3}. According to Ginzburg-Landau (GL) theory, $H_{\rm c2}$ can also be determined by using the formula : $H_{\rm c2} = H_{\rm c2}(0)(1-t^2)/(1+t^2)$, where $t = T/T_{\rm c}$ is the reduced temperature.  The solid line in Fig~\ref{fig3} gives a good fit to the above equation and $H_{\rm c2}$(0) value thus determined amounts to 27  and 20~T for $H~\parallel~$ [100] and [001] directions respectively.  These values are slightly larger than that  obtained from WHH theory. Using the expression $\xi$ = $(\Phi_0/2 \pi H_{\rm c2})$ for the  coherence length, where $\Phi_0$ = 2.07~$\times~10^{-7}~{\rm Oe~cm^2}$ we infer a coherence length of 40~\AA.

\begin{figure}[h]
\includegraphics[width=0.35\textwidth]{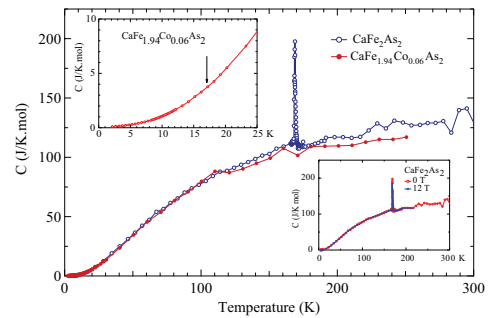}
\caption{\label{fig4}(Color online) Temperature dependence of the heat capacity of CaFe$_2$As$_2$ and CaFe$_{1.94}$Co$_{0.06}$As$_2$.  The high temperature SDW ordering is absent in the Co-doped sample. The lower inset shows the heat capacity of CaFe$_2$As$_2$ in 0 and 12~T field.  The upper  inset shows the temperature dependence of the heat capacity  of CaFe$_{1.94}$Co$_{0.06}$As$_2$ below 25~K.  }
\end{figure}

Fig.~\ref{fig4} shows the heat capacity from 1.8 to 300~K of CaFe$_2$As$_2$ and CaFe$_{1.94}$Co$_{0.06}$As$_2$. A sharp peak at 170~K indicates a first order like transition for CaFe$_2$As$_2$  similar to the previous reports~\cite{Ronning, Ni1}. However, in the Co-doped sample the peak at 170~K has vanished, which is in correspondence with the susceptibility and the resistivity data.  The heat capacity data varies smoothly across 90~K but at slightly higher temperatures one can imagine some kind of anomaly is present near 110~K.    The top inset of Fig.~\ref{fig4} shows the  the heat capacity of CaFe$_{1.94}$Co$_{0.06}$As$_2$ below 25~K.  The jump at $T_{\rm c}$ is not discernible from the heat capacity data.  Keeping in view the relatively broad superconducting transition as seen both in the susceptibility and resistivity, we believe the anomaly in the heat capacity at $T_{\rm c}$ is broadened and submerged in the background phonon contribution.  We may mention here that in Ba$_{0.55}$K$_{0.45}$Fe$_2$AS$_2$, where the superconducting transition is sharp, the anomaly in the heat capacity at $T_{\rm c}$ is barely discernible~\cite{Ni}. It may also be noted that the heat capacity of pure and Co-doped samples (main panel) is almost similar.  We infer that the Co-doping at $x$~=~0.06 level leave the phonon spectrum virtually unchanged.   

We have also measured the heat capacity of CaFe$_2$As$_2$ in a field of 12~T with the field direction parallel to both the crystallographic directions namely [100] and [001]  (lower inset of Fig.~\ref{fig4}).  The magnetic field has virtually  no effect on the SDW transition.  It has been pointed out that in a magnetic field the spin-density wave couples with bands of opposite spin leaving the nesting unaffected~\cite{McKenzie}.  

\begin{figure}[t]
\includegraphics[width=0.35\textwidth]{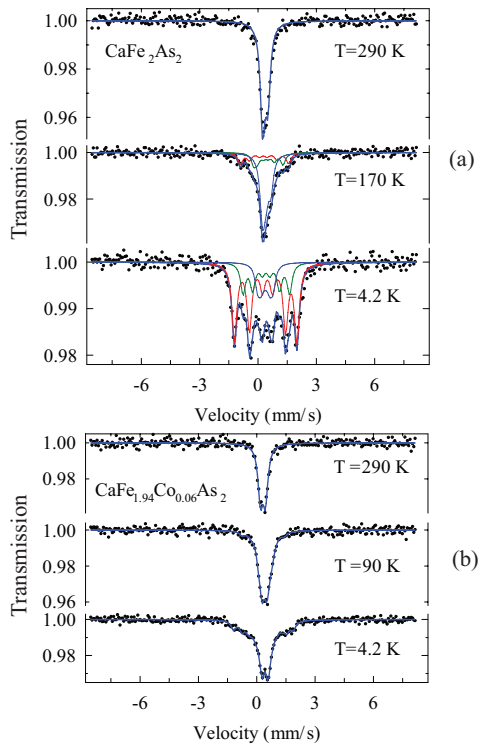}
\caption{\label{fig5}(Color online) M$\ddot{\rm o}$ssbauer  spectra of CaFe$_2$As$_2$ and CaFe$_{1.94}$Co$_{0.06}$As$_2$ at selected temperature.  For CaFe$_2$As$_2$ the three sub-spectra at 4.2~K are shown (see text)}
\end{figure}

The magnetic state of Fe in CaFe$_{2}$As$_{2}$ and CaFe$_{1.94}$Co$_{0.06}$As$_{2}$ was further investigated by recording  $^{57}$Fe M\"{o}ssbauer spectra at selected temperatures Fig.~\ref{fig5}. The room temperature spectra for both the samples were quadrupole split (eqQ/2=0.25mm/s) paramagnetic lines. In the pure compound the spectra are hyperfine field split below the SDW transition;  the spectrum at 170 and 4.2~K is fitted to a Hamiltonian consisting of magnetic and quadrupolar terms.  The best fit to the experimental spectrum was obtained by a superposition of  three sub-spectra with negligible quadrupole splitting ($<$0.1mm/s) corresponding to hyperfine fields of 10, 7.4 and 2.5~T with weightages of 60, 24 and 16~\%, respectively at 4.2~K.  

The occurrence of multiple sub-spectra with widely different hyperfine fields in the pure compound appears strange as there is only one Fe site in the crystal structure.  One reason could be the possible modulation of the magnetic structure incommensurate with the crystalline lattice, giving rise to  a hyperfine field distribution as reported in EuPdSb~\cite{Bonville}.  An incommensurate spin density wave order has been proposed as the cause of inhomogeneous broadening of the $^{75}$As NMR line in single crystal of BaFe$_2$As$_2$ grown from Sn flux, although the neutron scattering results find a commensurate order~\cite{Baek}.  On the other hand a commensurate SDW ordering has been reported on a self flux grown single crystal of BaFe$_2$As$_2$ from $^{75}$As NMR measurements~\cite{Kentaro}.  It has been claimed that the presence of even small amount of Sn in the single crystal can have a large effect on NMR spectra.  While our results indicate an incommensurate magnetic ordering, neutron diffraction on a single crystal of CaFe$_2$As$_2$ grown using Sn flux reveals a commensurate ordering~\cite{Goldman}.  We feel that the nature of SDW ordering needs further investigation. 

Turning our attention to the Co-doped sample, the purely quadrupolar split, paramagnetic  M\"{o}ssbauer spectrum  at room temperature persists down to about 100~K.  At 90~K an additional feature characterized by a broadening in the wings is observed which progressively increases down to  4.2~K.  Though it can arise due to relaxation effects the anomaly at 90~K in the magnetization data suggests the magnetic origin of the broadening.  The spectrum at 4.2~K can be simulated by a superposition of four subspectra corresponding to magnetic hyperfine fields of 9.5, 7.1 and 2.7~T with weightages of nearly 20\% each and a predominant paramagnetic component of 40~\%.  The M\"{o}ssbauer data thus corroborate  the weakening of the SDW with even a small amount of Co-doping, leading to the emergence of a superconducting state. This is consistent with the high pressure data~\cite{Park, Torikachvili2} where even a relatively low pressure of 2~kbar is sufficient to induce superconductivity with a $T_c$ of nearly 10 K.     

To conclude, we have grown large size single crystals of CaFe$_2$As$_2$ and CaFe$_{1.94}$Co$_{0.06}$As$_2$ by flux method using Sn as flux. Our $^{57}$Fe M\"{o}ssbauer results indicate that the SDW ordering in CaFe$_2$As$_2$ is incommensurate.  Co-doping at the Fe site reduces the unit cell volume marginally and it  weakens the SDW ordering and induces  superconductivity at a relatively high temperature of 17~K.  The upper critical field is estimated to be 20~T.

\end{document}